\documentclass[a4paper]{jpconf}
\usepackage{graphicx}
\begin{document}
\title{Density-Functional theory, finite-temperature classical maps,
and their implications for foundational studies of quantum systems.}

\author{M W C Dharma-wardana}

\address{National Research Council of Canada, Ottawa, Canada, K1A 0R6}

\ead{chandre.dharma-wardan@nrc-cnrc.gc.ca}

\begin{abstract}
The advent of the Hohenberg-Kohn theorem in 1964, its extension to finite-$T$,
 Kohn-Sham theory, and relativistic extensions provide
 the well-established formalism of density-functional theory (DFT).
 This theory enables the calculation of all static properties  of
 quantum systems {\it without} 
the need for an $n$-body wavefunction $\psi$. DFT uses the one-body density
distribution instead of $\psi$. The more recent time-dependent formulations
of DFT  attempt to  describe the time evolution of 
quantum systems without using the time-dependent wavefunction. 
 Although DFT has become the standard tool of condensed-matter 
computational quantum mechanics, its foundational implications have remained 
largely unexplored. While all systems require quantum mechanics (QM) at $T$=0,
the pair-distribution functions (PDFs) of such 
quantum systems have been accurately mapped into classical models at
effective finite-$T$, and using suitable non-local 
quantum potentials (e.g., to mimic Pauli exclusion effects).
 These approaches shed light on the quantum $\to$ hybrid $\to$ 
classical models, and provide a new way of looking at the existence of non-
local correlations without appealing to Bell's theorem. 
They also provide insights regarding
Bohmian mechanics. Furthermore, macroscopic systems even at 1 Kelvin have de 
Broglie wavelengths in the micro-femtometer range, thereby eliminating 
macroscopic cat states, and avoiding the need for {\it ad hoc} decoherence models. 
\end{abstract}

\section{Introduction.}
In 1964, Hohenberg and Kohn~\cite{dft1} proved a theorem asserting that the ground-
state properties of a stationary, non-relativistic system can be calculated
from a 
variational principle involving only the one-body density $n(\vec r)$ of the system, 
without recourse to the Schr\"{o}dinger equation and its wavefunction. 
The theorem was soon generalized to finite-$T$~\cite{Mermin}, and to
 relativistic systems~\cite{Rajagopal}. 
Thus it became clear that all thermodynamic properties of $n$-particle quantum 
systems, entangled,  interacting, or not, could be calculated without recourse to 
the $n$-body wavefunction. The method, re-written in the form of the Kohn-Sham 
theory has now become the preferred method in computational quantum mechanics (QM). 
Furthermore, since linear transport properties depend only on equilibrium 
correlation functions (c.f., the Kubo relations), a large class of quantum phenomena 
depends only on the density $n(r)$, a measurable quantity having a familiar  
`ontology' in quantum as well as classical mechanics. Unlike the many-electron
wavefunction of conventional QM, the one-body density is an observable. For
instance, the X-ray scattering from a specimen of matter gives direct
 information regarding  $n(r)$. The development of 
`time-dependent' density functional theory~\cite{tdft}, if substantiated,
 implies that 
density functional theory may enable a very different discussion 
of quantum phenomena than has been the custom in foundational studies.
 In this article, for  the sake of brevity and simplicity, we limit ourselves
 mainly to non-relativistic equilibrium systems. In the following presentation
we look at DFT and compare it with relevant aspects of Bohmian mechanics, 
treatment of hybrid systems within DFT, as well as  classical
 representations inspired by DFT ideas. Furthermore, finite-$T$ physics
 leads us to conclude that there is no need for {\em ad hoc} decoherence
 theories to eliminate macroscopic cat states from QM.


\section{Density functionals and Bohmian mechanics.}
To be specific, we consider a system of electrons subject to an external
potential $v_{ext}(\vec{r})$, and interacting with one another via the
 Coulomb potential. In many cases this external potential is produced by a fixed set
of nuclei, as in molecular physics or quantum chemistry calculations at $T=0$.
However, the ions cannot be treated as fixed in dealing with liquid metals
and hot plasmas.
Hence, in discussing true hybrid systems, we allow the nuclei (or ions) to be
 a second interacting subsystem, interacting with the electron subsystem, and the
 finite-$T$ discussion becomes not only appropriate, but also necessary.
However, in simplified models, e.g., the jellium model, the ions provide a
 fixed uniform positive distribution of charge that exactly neutralizes the
negative charge of the electron distribution. The $q\to 0$ singularities
 in the Coulomb potential $4\pi/q^2$ for electron-electron or ion-ion interactions
are exactly canceled by the corresponding term in the ion-electron interactions.

However, while keeping such specific models in mind, we can simply work with
an unspecified external potential $v_{ext}$ and a corresponding one-body
 electron distribution $n(r)$. Then according to Hohenberg and
 Kohn~\cite{dft1}, the exact
 ground state density $n(r)$ and the ground-state energy $E_v[n]$ corresponding
 to that external potential $v_{ext}$ are obtained by
 the minimization of the following energy functional.
\begin{equation}
\label{eq-dft1}
E_v[n]=T[n]+U_{cou}[n]+\int d\vec{r}v_{ext}(\vec r)n(\vec r)
\end{equation}
The kinetic energy contribution, viz., $T[n]$, and the Coulomb electrostatic energy
$U_{coul}[n]$ are said to be `universal functionals' of $n(r)$, in the sense that
they do not depend on the external potential $v_{ext}(\vec r)$ specific to the 
system. The original Hohenberg-Kohn theorem was established for non-degenerate 
ground states at $T=0$. This has since been restated within the language of 
functional analysis~\cite{Lieb82}, while degenerate ground states, spin-polarized 
systems, finite-$T$ systems, relativistic systems, multi-component hybrid systems,
superconductors, liquids and plasmas etc., have been studied within the
first three decades of DFT~\cite{Ilciacco}.

It is clear that a direct minimization of $E_v[n]$ via
 $\delta E_v/\delta n(\vec(r)$
would require a knowledge of the kinetic energy functional $T[n]$, and the
 electrostatic potential $U_{cou}[n]$. The latter is a known functional of
 $n(r)$, but $T[n]$ is unknown. Although this is a universal functional of
 $n(\vec r)$, it turns out to be highly non-local. Hence
gradient expansions merely lead to Thomas-Fermi like theories. These,
sometimes known as `orbital-free DFT'~\cite{KaraiTrickey12}, presently lack the full
accuracy of methods that use a wavefunction.
Since DFT is presented via a one-body density, the Kohn-Sham formulation
of DFT uses a one-body `Kohn-Sham wavefunction' to construct a
 non-local kinetic energy, where a single Kohn-Sham electron moves in
 the electrostatic potential of the other electrons and an additional
 exchange-correlation potential $V_{xc}[n]$
which is a functional of the one-body density. Thus, the many-electron problem 
is replaced by an effective one electron problem in Kohn-Sham theory. Here the
Kohn-Sham electron moves in a non-local Kohn-Sham potential that is made up
of the electrostatic potential $U_{cou}[n]$, and the non-local
 $V_{xc}[n]$. However, unlike $T[n]$, approximate models of
 $V_{xc}[n]$ have proved to be quite successful, when used in
 the Kohn-Sham equation where $T[n]$ is constructed for each case,
instead of relying on a knowledge of the universal functional $T[n]$.    

It is of interest at this stage to examine Bohmian mechanics, where the
 Schr\"{o}dinger many-particle wavefunction $\psi(\vec x_1,\cdots \vec x_n)$
 is used to construct a non-linear equation of
motion containing the external potential as well as a quantum potential 
$Q(\vec x_1,\cdots \vec x_n)$. Let us limit ourselves to a single particle
problem in 1-D for the sake of brevity.
 Bohm~\cite{Bohm1952} writes the wavefunction in the form:
 \begin{eqnarray}
\label{eq:psi_RexpS}
 \phi(x,t)&=&R(x,t) \exp\{iS(x,t)/\hbar\} \\
 R(x,t)&=&\surd{n(x,t)};\;\; n(x,t)=|\phi(x,t)|^2
 \end{eqnarray}
 so that $R$  (positive), and $S$ are real functions. Substituting
 this form of the
 wavefunction into the Schr\"{o}dinger equation, one obtains Bohm's
 form of the equations as:
 \begin{eqnarray}
 \label{bo1}
 \frac{\partial n}{\partial t}+\nabla . \left[n \frac{\nabla S}{m}\right]&=&0 \\
 \label{bo2}
 \frac{\partial S}{\partial t}+v_{ext}+\frac{(\nabla S)^2}{2m}+Q(x,t)&=&0\\
\label{qpot-eq}
Q(x,t)&=&\frac{-\hbar^2}{2m}\frac{\nabla^2R}{R}
 \end{eqnarray}
Eq.~\ref{bo1} is simply the equation of continuity. The term $(\nabla S)^2/2m$
in Eq.~\ref{bo2}  \index{equation of continuity}
is interpreted as the kinetic
energy of a point-like particles with momentum $p$ = $\nabla S$. In effect,
 Eq.~\ref{bo2}
has the form:
\begin{equation}
\label{eq:bohm-HamJ}
\partial S/\partial t+H(p,x)+Q(x,t)=0
\end{equation}
This is exactly like a classical Hamilton-Jacobi equation
 with the extra potential $Q(x,t)$, called the
quantum potential,  or the Bohm potential. Since we are interested in
 the simplest form of DFT (systems in equilibrium) in this presentation,
 let us consider an electron
confined within an infinite potential well of width $a$. The eigenfunctions
are of the form $\psi_n(x)=\surd(2/a)\sin(k_n x);\;\;\; n=1,2,\cdots$. It is
easily shown that the particle current $j(x,t)$ is zero, and the Bohmian
particle is {\it at rest} in some unknown location $x$ in the well, with the
probability $|\psi_n(x)|^2$ of {\it being} there. Einstein felt that the
particle being at rest in the well was not the correct physical 
picture. A more detailed discussion
of the Bohmian particle in a well, and a possible resolution to
 Einstein's objections~\cite{Ein-festschrift}  are given in
 Chapter 6 of Ref.~\cite{WS-bk}. 

Does the Bohmian particle in the well have no kinetic energy? It is easy
 to show that the kinetic energy is completely resident in Bohm's quantum
 potential while the particle current is zero. In fact, the kinetic energy
 functional $T[n(x)]$ per electron confined in the quantum well, given by
 Hohenberg-Kohn theory (or conventional QM) is essentially the quantum
 potential of Bohm. Although some advocates of Bohmian mechanics have moved
away from the quantum potential (mainly due to philosophical reasons), we
find it to be of a sort similar to the effective potentials based
on density-functionals used in
 condensed-matter many-body physics. 

For instance, one of the objections  adduced against
 the quantum potential is that
it has the `unusual effect' of making the particle
 density $n$ influence the
dynamics of individual particles. However, that is precisely the
message of DFT. It is typical of many-particle physics that
the individual dynamics is influence by the over-all particle
distribution. For instance, a low-order (mean-field) many-body
theory like the Hartree theory of many electrons requires that an
electron moves in the many-body field of the other electrons
produced by the density distribution $n(x)$. Hartree-Fock theory
involves a highly non-local Hartree-Fock potential dependent on the
density $n(r)$.

Interestingly, D\"{u}rr, Goldstein and Zhangi~\cite{DGZ95}
 summarize their attitude
to the quantum potential with the statement, `We believe that the most
 serious flaw in the quantum-potential formulation of Bohmian mechanics is
that it gives a completely wrong impression of the lengths to which
we must go in order to convert orthodox quantum theory into something
more rational. The quantum potential suggests \ldots we must incorporate
 into the theory a quantum potential of a {\it grossly} non-local character'.
We emphasized the word `grossly' in the above to high-light where the 
attitudinal squeamishness resides. In contrast, most physicists who use the
highly non-local  Hartree-Fock potential as a basic first step
 in their calculations have given up any such squeamishness.
 Regarding `the lengths'
 needed to reach the
classical Hamilton-Jacobi form, as seen from above,
 we need only a few lines.

 The insights from DFT enable us to consider the Bohm potential as
 a functional of the density, without invoking a wavefunction.
 Furthermore, Eq.~\ref{qpot-eq} may be used as a differential equation
 for the direct determination of $Q$  given in terms of $R=\surd{n}$,
 subject to satisfying the equation of continuity, and minimizing the energy.
Once $R$ is determined by iteratively solving Eq.~\ref{qpot-eq},
 $S$ can be obtained from the continuity equation. In effect,
a time-dependent system is described by the four vector consisting of the
current $j(x,t)$ and the density $n(x,t)$. These are precisely the quantities
treated in DFT, and in Bohmian mechanics.

The above considerations suggest that Bohmian mechanics can be
 formulated entirely in a language free of the wavefunction, using only
 density functionals, exactly as in DFT.
\section{Pair-distributions and classical maps.}
\label{gr-sec}
The Hohenberg-Kohn theorem reduces the $n$-electron problem to that of
a functional of an effective one-body problem and an exchange-correlation
 potential $V_{xc}[n]$. Given that the basic electron-electron interactions
 (that lead to quantum correlations including those of the Bell type) are
 pair-interactions and Pauli exclusion effects, the reduction to a one-body
 problem achieved in DFT appears a bit mysterious, and one would expect
 that fundamental two-particle correlations are embedded in the $V_{xc}$. 
In fact, the latter is most transparently evaluated as a coupling constant
 integration over the pair-distribution function (PDF), i.e.,
  $g(\vec r_1, \vec r_2)$ of the system.  This is the probability of
 finding an electron at $r_2$, given an electron at $r_1$.    

The PDF is well-defined both classically and quantum mechanically. In QM, the
PDF is calculated from the many-particle wavefunction by integrating over
 all except two space variables $\vec r_1,\vec r_2$ of two electrons. In
 a classical system, we can consider  a uniform electron fluid (as in jellium),
 with a mean density $n$ and Wigner-Seitz radius $r_s=\{3/(4\pi n)\}^{1/3}$.
We position the origin on the first electron, so that $\vec r_1=0$.  Then,
 invoking radial symmetry and uniformity, we can use $g(r)$ rather than
 $g(\vec r)$ for the PDF, where $r$ is the radial distance of the (second)
 electron from the origin. When classical mechanics holds (i.e., for $T> 0$
 and for very small de Broglie wavelengths), this $g(r)$ for an interacting
 Coulomb fluid can be calculated using the modified-hyper-netted-chain (MHNC)
 equation~\cite{HansenMac}, or via molecular dynamics (MD).

In QM, even the non-interacting electron system is of interest if all the
 particles have the same spin. Then the Pauli exclusion principle controls the
entanglement of the  $n$-electron system. 
This non-interacting set of $n$-entangled electrons is described by a
 Slater determinantal wavefunction. The fully spin-polarized system
 (parallel spins $\sigma$), and the fully unpolarized system (anti-parallel
 spins $\sigma_1\ne\sigma_2$) are of interest. They can be analytically
 treated at $T=0$ as in Ref.~\cite{MahanBk}, and extended to finite-$T$
 as in Ref.~\cite{prl1}. Furthermore, the Pauli exchange effect (which leads
 to entanglement since the wavefunction is a Slater determinant) can be
 re-written as a non-local quantum potential which
 is a universal function of $r/r_s$.  This `Pauli-exclusion' potential
 $P(r/r_s)$ can be constructed such that it reproduces the quantum $g(r)$
  when used in the classical MHNC equation. This provides an interesting
 and useful classical map of the quantum problem that has been very fruitful
 in applications to interacting quantum problems
ranging from 2-D electron layers~\cite{prl2}, hydrogen plasmas, electrons
 in graphene, and quantum dots~\cite{sannibel}. That is, correlated,
 interacting quantum problems
 can be replaced by equivalent classical problems where pair-potentials
 involve additional potentials containing quantum effects which are
 usually non-local. Bohmian mechanics achieves this directly from the
 $R=\surd(n)$ part of the wavefunction. DFT achieves this via functionals
 of the one-body density $n(r)$. Naturally,
 non-local contributions are usually quite important in both DFT and
 Bohmian mechanics.
\begin{figure}[t]
\center
\includegraphics[height=6 cm, clip]{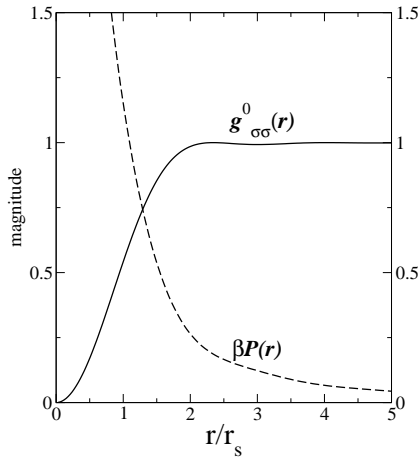}
\caption{The pair-distribution function $g^0_{\sigma\sigma}(r)$ 
 gives the probability of
finding a $\sigma$-spin electron at $r$ if there is already
 a $\sigma$-spin electron at the
origin  (Coulomb repulsion neglected). 
We plot $r/r_s$ on the $x$ axis, and the result is valid
for any density, $r_s$ being the sphere-radius containing an average of
 one electron. Thus
the `entanglement' due to Pauli exclusion holds even if $r_s$ is a 
lunar distance.  The Pauli exclusion is exactly
mimicked by a scale-independent
repulsive potential $\beta P(r)$ when used in a classical MHNC equation. 
 }
\label{pdf-fig}
\end{figure}

The PDF of non-interacting fully spin-polarized entangled electrons,
 denoted by $g_{\sigma, \sigma}^0(r)$ is shown in Fig.~\ref{pdf-fig}.
It has the form:
\begin{eqnarray}
\label{eq:gr0}
g^0_{\sigma\sigma'}(r)&=&1;\;\;\;\sigma\ne\sigma' \\
g^0_{\sigma\sigma}(r)&=&1-\{3j_1(rk_F)/(rk_F)\}^2
\end{eqnarray}
(The large-$r$ behavior is normalized to unity, whereas
if $\zeta=0$, it is natural
 to normalize the spin-resolved functions to 1/2).
Here $j_1(x)$ is the spherical  Bessel function, while
its argument $x$ contains $k_F=1/(\alpha r_s), \alpha=0.52106$,
where $k_F$ is the Fermi momentum.
 The probability of finding an
electron at $r$, given that there is an electron at the origin of coordinates
 is affected by their
mutual Coulomb interaction. This is not included in
 the above calculation, and hence
the superfix `zero' in $g^0$. A treatment including the Coulomb
potential is given in Ref.~\cite{prl1}, where it is shown that
remarkably good agreement with Monte-Carlo quantum simulations
 is obtained for these classical maps.

 Since the PDF scales as $r/r_s$, the correlations remain undiminished,
 whatever be the mean-separation $R=2r_s$ that is imposed on any two
 electrons in the entangled system. That is, {\it local realism} as
 required by the Einstein-Podolsky-Rosen gedanken experiment does
 not hold. Thus the correlations unearthed by the Bell inequalities
 are very transparently and simply revealed by an examination of
 the PDF of correlated electron systems. 
This topic is discussed in greater detail in Chapters 7 and 8 of
 Ref.~\cite{WS-bk}.
\section{Finite temperatures, hybrid systems and decoherence.}
\label{finteT-sec}
Formal discussions of hybrid systems (i.e., systems containing quantum
particles as well as classical particles)~\cite{Heslot85} rarely
 bother to discuss the
role of temperature, partly because some of these discussions
deal with model problems dealing with just one or two particles.
In reality, all particles in any system obey quantum mechanics.
 However, if some
particles have de Broglie wavelengths $\lambda$ smaller than
some assigned length scales associated with the system, then such
particles among themselves can be treated within classical mechanics.
The interactions between the particle deemed to be classical, and 
those deemed to be quantum occur via a Hamiltonian $H_{int}$ whose
characteristic length scales and energy scales have to be
 investigated separately. In the following we look at systems
at finite temperature $T$.

If a system is coupled to a heat bath and held at some finite
 temperature $T$,
the average kinetic energy of the particles is a function of $T$.
 In the classical limit the average kinetic energy of an atom
 is $3T/2$, using energy units for $T$
where the Boltzmann constant becomes unity. At $T=0$, for our model
 system of fermions (jellium), the kinetic energy becomes $3E_F/5$,
 where $E_F$ is the Fermi energy. The latter can be thought of as
 a temperature $T=2E_f/5$. Hence
we see that we can ascribe a de Broglie wavelength $\lambda$ and 
a thermal momentum $mv$ to any particle in a system at an effective
 temperature $T$, with $v$ and $\lambda$ given by:
\begin{equation}
\label{deBrog-eq}
v=\sqrt{3T/m}, \;\;\; \lambda=h/p=h/\sqrt{3mT}
\end{equation}
It is clear that $\lambda$ is infinite for systems at $T=0$, or for
sufficiently dilute quantum systems ($E_F\to 0$ as $n \to 0$). Thus all
systems become quantum systems as $T\to 0$  since their de Broglie wavelengths
become infinite. However, if at some finite (even if low) temperature,
 e.g., 1 Kelvin, if the mass of a particle is sufficiently large,
 then if $\lambda \ll 1$ atomic unit, we may assume that superposition
 and entanglement become unimportant. Such
heavy particles are classical particles. Hence, since all systems
 are quantum systems at $T=0$, and heavy particles become classical
 at some finite-$T$, any quantum theory of hybrid systems~\cite{Heslot85}
 has to ultimately incorporate a quantum treatment of
thermal fields. 

The conventional quantum approach to finite-$T$ problems is
difficult, and involves a typical doubling of the Hilbert space as in
 Umezawa's thermofield dynamics~\cite{Umezawa},   or in the
 Martin-Schwinger-Kedysh contour technique~\cite{Martin-Schwinger}.
 These methods are useful for tackling a class of non-equilibrium physics.
  However, even
static finite-$T$ quantum calculations (e.g., via the Matsubara
 method~\cite{MahanBk}) are restricted to perturbation theory.

On the other hand, finite-$T$ density functional theory~\cite{Mermin} presents
 a computationally and conceptually much simpler approach to
 finite-$T$ problems, as well as two-temperature quasi-equilibria.
 Here the finite $T$ one-body density functionals $n(\vec x, T)$ determine
 all the thermodynamics and linear transport properties of the system
{\it without} the need for a wavefunction.

The explicit construction of a DFT and the calculation of such
 finite-$T$ hybrid systems for a system of protons (classical) and
 electrons (quantum) at finite-$T$ has been given in Ref.~\cite{hyd0}.
 The method is applicable to any electron-ion system, where the ions are
 treated classically. The usual Born-Oppenheimer approximation is
 not needed. If the ion distribution is denoted by $\rho(\vec r)$,
 and the electron distribution is $n(\vec r)$, the total free
 energy $F([\rho], [n])$ is a functional of both distributions. Hence we have
two coupled Hohenberg-Kohn-Mermin type equations to determine the
 finite-$T$ distributions of the hybrid system. That is, the functional
 derivatives with respect to variations $\delta n, \delta \rho$ lead
 to two coupled Euler-Lagrange equations.
\begin{equation}
\label{dft-hybrid-eq}
\frac{\delta F[\rho, n]}{\delta n}=0,\;\; \frac{\delta F[\rho, n]}{\delta \rho}=0.
\end{equation}
The first of these equation can be reduced to a Kohn-Sham equation
 for electrons at
 finite-$T$, or simply retained as a Hohenberg-Kohn variational form.
 The second equation leads to a classical Hohenberg-Kohn equation for the ion
distribution that is identified as a form of the MHNC equation. In effect,
 it is a Boltzmann distribution for the density $\rho$ in terms of the
 `potential of mean force' (PMF) used in the statistical mechanics
of liquids~\cite{HansenMac}. The PMF is the appropriate Kohn-Sham potential
 for this classical system.
The full practical implementation of this approach to a typical hybrid
 system (an interacting gas of electrons and protons at finite $T$
 inclusive of the formation of hydrogenic bound states)  is given
 in Ref.~\cite{hyd0}.
\subsection{Decoherence and macroscopic cat states.}
\label{cat-lambda-eq}
The equation \ref{deBrog-eq} is of great importance in showing us that
 there is no need for {\it ad hoc} theories of decoherence to
 eliminate macroscopic cat states from QM. Let us consider a
 1 kg cat at room temperature (300 K, i.e., $9.500\times  10^{-4}$ Hartees).
 The atoms in the cat have internal energies with approximately  $3T/2$ arising
 from translational motion. The center of mass will also make random
 oscillatory motions about an equilibrium position consistent with
 its temperature and any motion of the cat. The corresponding
 center-of-mass de Broglie
 wavelength $\lambda_{c}$ turns
 out to be 
\begin{equation}
\label{lambda-cm-cat.eq} 
\lambda_{c}\simeq 9.45\times 10^{-23}\;\mbox{meters}
\end{equation}
The `radius' of a proton is about 0.88 femtometers
(1 fm=10$^{-15}$ meters). The de Borglie wavelength  of the cat
 is about one hundred-millionth of the size 
of a proton! Even when cooled to one Kelvin,
 the de Broglie wavelength
 remains totally negligible.
Even a free electron close to a cat's
body would not get entangled with the cat's quantum energy states.
 Matter has to
be squashed to densities where nuclear reactions begin before
superpositions are possible at such small values of $\lambda_c$.
Clearly, Schr\"{o}dinger's cat states of macroscopic bodies
(entanglements or superpositions) do not exist in nature
 except at $T$=0, or possibly in white dwarfs.

One may argue that the mass $M$ of the center of mass 
is not the proper quantity to use. 
One may claim that QM should be applied to some molecules of the cat that
are specific to it being dead or alive in Schr\"{o}dinger's
 cat paradox. In that case we are no longer
applying QM to a macroscopic system and then there is no difficulty in
having superpositions or entangled states. Chemical reactions
 (e.g., cyanide reacting with the cat) are simply processes where
 reactive atomic groups enter into entanglements. Such processes
 pose no difficulty and are not `paradoxical'.

One may also question 
the use of a momentum calculated from $3T/2$. This is the classical
kinetic energy of an ideal gas. We know from Dulong and Petit's
law (or from the more modern Einstein-Debye theory of solids) that
classically, a system held together by chemical bonds, e.g., a cat, 
has three degrees of vibration per atom, and three degrees of
 translation, each having an energy $T/2$. For solids in the low-temperature
limit, we need the Fermi energy of the system. Such detailed
treatments merely introduce numerical factors of little consequence. The
basic conclusion that $\lambda_c$ is very very small remains firm.   

These considerations imply that explicit decoherence theories like those
of Penrose (`objective reduction' due to quantum gravity)~\cite{Penrose}, or
the `spontaneous localization' model of Ghirardi, Rimini and 
Weber~\cite{GRW1986} are unnecessary additions to quantum mechanics,
 if we grant that the relevant de Broglie lengths are much smaller
 than even nuclear
radii.  
\section{conclusion.}
The tradition of studying just one or two quantum particles, usually in
 a two-state model has a long history in foundational studies. However,
 if the many-body problem is examined, QM can be approached using one-body
 density functionals, without recourse to wavefunctions, both at $T=0$
 and at finite-$T$. An examination of Bohmian mechanics within the
 perspective of DFT is conceptually rewarding and clarifies the
 nature of the Bohmian quantum potential. Finite-$T$ quantum mechanics
 leads us to consider the thermal de Broglie wavelength which, being
 negligible for macroscopic systems, ensures that there are no macroscopic
 Schr\"{o}dinger cat states.

\end{document}